\documentclass[aps,twocolumn,prb,tightenlines,floatfix,superscriptaddress]{revtex4-1}

\usepackage{graphicx}
\usepackage[english]{babel}
\usepackage{amsmath}
\usepackage{amssymb}
\usepackage{slashed}
\usepackage{pdfpages}

\newcommand{\mean}[1]{\langle #1 \rangle}

\def\q{{\bf q}}
\def\k{{\bf k}}

\def\w{\omega}

\allowdisplaybreaks[4]

\newcommand{\mrm}[1]{{\mathrm{#1}}}

\begin{document}

\title{ Unified treatment of Fermi pockets and arcs scenarios for the cuprates:
Sum rule consistent response functions of the pseudogap}
\author{Peter Scherpelz}
\author{Adam Ran\c{c}on}
\affiliation{James Franck Institute,
University of Chicago, Chicago, Illinois 60637, USA}
\author{Yan He}
\affiliation{College of Physical Science and Technology, Sichuan University,
Chengdu, Sichuan 610064, China}
\author{K. Levin}
\affiliation{James Franck Institute,
University of Chicago, Chicago, Illinois 60637, USA}

\date{\today}


\begin{abstract}
Essential to understanding the cuprate pseudogap phase is
a study of the charge (and spin) response functions, which we address here
via a consistent approach to the
Fermi arcs and the Fermi pockets scenario of Yang, Rice and Zhang (YRZ). The
two schemes are demonstrated to be formally similar, 
and to share a common physics platform; we use this consolidation to
address the inclusion of vertex corrections which have been omitted in YRZ applications.
We show vertex corrections can be easily implemented in
a fashion analytically consistent with sum rules and that they
yield important contributions to most observables.
A study of the charge ordering susceptibility of the YRZ scenario makes their simple
physics evident:
they 
represent the inclusion of charged bosonic, spin singlet degrees
of freedom, and are found to lead to a double peak structure. 
\end{abstract}
\maketitle

The discovery of the high temperature superconductors has led to the development
of extensions (as well as replacements) for BCS theory in which strong
correlations or self energy effects are present simultaneously with the
underlying pairing interactions which drive superconductivity.  These
self energy contributions are associated with the anomalous
pseudogap behavior which sets in above $T_c$ and which may persist as well below
the transition. A proper treatment of highly correlated normal and superconducting states
introduces consistency constraints (vertex corrections, Ward identities and
sum rules). This was central in
the long history of BCS theory, where these constraints  
led to an
understanding of new types of ``particles" or excitations such as the Higgs
boson and its related mechanism.

In the high-$T_c$ cuprates, characterizing the spin and charge response has
been essential for clarifying whether the pseudogap is associated with pairing
or with an alternative ordering, although there is as yet no unanimity.\cite{LeeReview,ourreview} A growing enthusiasm is emerging for one
particular pairing-based approach to the pseudogap developed by Yang, Rice and
Zhang (YRZ),\cite{YRZ} which suggests the possibility of charge ordering
in the presence of pairing.\cite{Comin} Accompanying this interest has been a fairly
universal neglect of vertex corrections\cite{Bubble,Comin} in the calculated
response functions.  This omission is
not a formal technicality.  At a minimum such corrections are essential in order
to ensure that the normal phase is not associated with an unphysical Meissner
effect.

This leads to the goal of the present paper, which is to present a calculation
of self consistent response functions for the YRZ theory of Fermi pockets\cite{YRZ} along
with an alternative approach
involving Fermi arcs.\cite{Chen2,ourreview,Chen4} We show here that
these two approaches to the pseudogap are in fact
closely related, sharing common
physical features and allowing nearly identical calculations of vertex
corrections. We also show
that these vertex corrections are consistent with sum rule constraints.
Finally, we
demonstrate
that introducing self consistency leads to
(hitherto ignored\cite{Bubble,Comin}) contributions to the spin and
charge response, which are of sizable magnitude and can be physically understood. 

The consolidation that we present between 
Fermi arcs and pockets is possible because
both theories contain pairs which are present in the pseudogap phase. These
pairs, with their bosonic character, lead to similar vertex corrections in both
theories. Formally, these pseudogap pairs arise from the semi-microscopic
self energies posited by the theories,\cite{ourreview,YRZ} 
which contain both superconducting (sc) and
pseudogap (pg) components: $\Sigma = \Sigma_\mrm{sc} + \Sigma_\mrm{pg}$. 
The form of $\Sigma_\mrm{pg}$ is rather similar to the BCS-like
self energy of the condensate but in the pockets case this
term leads to a reconstructed Fermi surface (``pockets")
and in the arcs case to a blurring of the $d$-wave nodes
(``arcs").
This two-gap form of $\Sigma$
ensures that the pseudogap correlations
persist below $T_c$, but are
distinct from condensation.
It should not be confused with (one-gap) phase
fluctuation models, where it is presumed that the pseudogap turns into a
condensate gap at the transition. 
We show here how to impose consistency for both two-gap approaches
by addressing the f-sum rule on the charge density response (above
$T_c$) and the transverse sum rule on the current density
response at all $T$.
In this way
vertex corrections emerge naturally
and can be readily incorporated into the Fermi pockets approach of YRZ.
(They have been included in the formally related Fermi arcs approach in 
Refs.~\onlinecite{Chen2,Kosztin2}.)

For the pockets model of YRZ the microscopic picture for the pg contribution is
that it is associated with resonating pairs of spin
singlets\cite{RVB,Andersonbook} which, when holes are injected, become charged.
In the Fermi arcs model, where we consider a two-gap rendition,\cite{ourreview}
(which
introduces both sc and pg gaps $\Delta_\mrm{sc}$ and $\Delta_\mrm{pg}$, as in
YRZ), the pg
correlations represent finite momentum, out of the condensate excitations; they
reflect a stronger-than-BCS attractive interaction.  This scenario for a
pseudogap is realized in the laboratory of ultracold Fermi gases
\cite{ourreview} and associated with BCS-BEC crossover.  The excited pairs are
\textit{gradually} converted to condensed pairs as the temperature is lowered
below $T_c$. Here $\Delta_\mrm{pg}^2$ is effectively zero at temperature $T=0$
and reaches a maximum at $T_c$; in this way the square of the excitation gap
$\Delta^2 = \Delta_\mrm{sc}^2 + \Delta_\mrm{pg}^2$ is relatively constant below
$T_c$.  Just as in the YRZ pockets model, this Fermi arcs model has addressed
thermodynamics\cite{Chen4}, Nernst\cite{Tan}, the penetration
 depth\cite{Chen2,JS1}, quasiparticle interference in STM\cite{ourqpi,ourqpi2} as
well as the ac and dc conductivities\cite{ouroptical,ourthz,Ourconduct} and
diamagnetism.\cite{ourdia}  Perhaps its greatest success is that it naturally
leads to a nodal-antinodal dichotomy.\cite{ourarpes}  This refers to the
collapse of the arcs as temperature approaches $T_c$ from above; as $T$ 
approaches
$T_c$ from below the nodal ARPES gap has a $T$ dependence which reflects that of
the order parameter, $\Delta_\mrm{sc}$, while the antinodal gap is very little
affected by the transition.
 
\emph{Theory and response functions.} We introduce the Green's function (and neglect for simplicity the incoherent
contributions) 
\begin{equation}
G_K = \frac{1}{\omega - \xi_\k - \frac{\Delta_\mrm{pg}^2}{\omega + \xi^\mrm{pg}_\k}
-\frac{\Delta_{sc}^2 }{\omega +\xi_\k + \Sigma_R (\k,-\omega)}},
\end{equation}
where $K=(\w,\k)$. Here for the arcs and pockets models respectively
\begin{alignat}{5}
\xi^\mrm{pg}_\k &= \xi_\k + i \gamma \quad
&& \mrm{and} \quad & \Sigma_R (\k,\omega) &\equiv 0,
\notag \\ 
\xi^\mrm{pg}_\k &= \xi^0_\k \quad && \mrm{and} \quad & \Sigma_R(\k, \omega) &= 
\frac{\Delta_\mrm{pg}^2}{ \omega+ \xi^\mrm{pg}_\k},
\end{alignat}
where the dispersion $\xi^{(0)}_\k$ is introduced in Ref.~\onlinecite{YRZ}.
There are two different assumed forms\cite{RiceReview}
for the sc piece in the YRZ approach, and here we take
the original one,\cite{YRZ} rather than introduce corrections associated with
phenomenological adjustments.  Similarly
we stress that for the arcs model we can minimize phenomenological input and simply take
the central free parameter $\gamma$ as independent of temperature.
The role of $\gamma$, which has a microscopic
basis,\cite{Malypapers} is
critical; it leads to a smearing of the $d$-wave node and thus to the Fermi 
arcs.\cite{NormanArcs,FermiArcs,ourarpes,Chubukov2}

The pseudogap and superconducting self energy in both schemes are given by
\begin{eqnarray}
\Sigma_\mrm{pg}(K) &=& - \Delta_\mrm{pg}^2 G_0^\mrm{pg}(-K) = 
-\Delta_\mrm{pg}^2 
\times \frac{1}{\omega + \xi^\mrm{pg}_\k }, \nonumber \\ 
\Sigma_\mrm{sc}(K) &=&
-\Delta_\mrm{sc}^2 G_0^\mrm{sc}(-K)
=
-\Delta_\mrm{sc}^2 \times \frac{1}{\omega +\xi_\k + \Sigma_R (\k,-\omega)}
\nonumber 
\end{eqnarray}
which defines $G_0^\mrm{pg}$ and $G_0^\mrm{sc}$.
Because $\xi^\mrm{pg}_\k \neq \xi_\k $, the YRZ scheme arrives at a
many-body reconstructed bandstructure.
Moreover, we see from $G$ in both the arcs and pockets models
that the form of $\Sigma_\mrm{pg}$ is not very different from that
of $\Sigma_\mrm{sc}$, yet their effects on the physics of the generalized
response functions have to be profoundly different.  
We enforce this difference by ensuring that
there can be no Meissner effect in the normal phase, and this
requires the inclusion of vertex corrections in the current-current
response function which we write as $\tensor{P}$. 
It
will be convenient to introduce a parameter
$\Lambda^\mrm{sc} \equiv 1$ for the pockets case and 
$\Lambda^\mrm{sc} \equiv 0$ for the
arcs scenario. We also define
\begin{eqnarray}
F_{\mrm{pg},K}
&\equiv& - \Delta_\mrm{pg} G_0^\mrm{pg}(-K) G_K ,\nonumber \\ 
\label{eq:3} 
F_{\mrm{sc},K} &\equiv& - \Delta_\mrm{sc} G_0^\mrm{sc}(-K) G_K.
\end{eqnarray}
The quantity $F_\mrm{pg}$ (unlike $F_\mrm{sc}$) is \textit{not} to be associated with
superfluidity.
It is not in the notation ``$F$" that superfluidity enters, it is in the way in which the
current-current correlator is constructed, as we show below.

Next, we obtain an expression for the diamagnetic current contribution
$\big(\frac{n}{m}\big)_\mrm{dia} \equiv 
2\sum_K\frac{\partial^2\xi_{\mathbf{k}}}{\partial\mathbf{k}\partial\mathbf{k}}G(K).
$
For
notational simplicity we drop terms which involve the $k$ derivative of the $d$-wave
form factor throughout.
These effects
can be readily inserted, but are seen to be negligible in magnitude. We find
that the diamagnetic current can be rewritten
via integration by parts, 
using $\partial G (K) /\partial \k = - G^2(K) 
\partial G^{-1}(K) /\partial \k$ so that
$\big(\frac{n}{m}\big)_\mrm{dia}$
\begin{eqnarray}
&=& 
-2\sum_KG^2_K\frac{\partial\xi_\k}{\partial \k}
\frac{\partial\xi_\k}{\partial \k}
+
2\sum_KF_{\mrm{pg},K}^2 \frac{\partial\xi_\k^\mrm{pg}}{\partial \k}
\frac{\partial\xi_\k}{\partial \k} \label{eq:7} \\  
&+& 
2\sum_KF_{\mrm{sc},K}^2 \big[ 
\frac{\partial\xi_\k}{\partial \k}
\frac{\partial\xi_\k}{\partial \k}
-
\Lambda^\mrm{sc} \frac{\partial\xi_\k^\mrm{pg}}{\partial \k}
\frac{\partial\xi_\k}{\partial \k} \Delta_\mrm{pg}^2 (G_0^\mrm{pg}(K))^2
\big] .\nonumber 
\end{eqnarray}
Given the parameterized self energies introduced above, \textit{in this exact expression, central to this paper, the second 
term on each line provides a template for the form of
the ignored vertex corrections in
the response functions}.
That there is
no Meissner effect above $T_c$ implies
that the current-current correlation function at zero wavevector
and frequency, 
$\tensor{P}(0)= - \big(\frac{n}{m}\big)_\mrm{dia}$.  
Below $T_c$ in the YRZ scheme we  
make use of the  superconducting
Ward identity\cite{Schrieffersbook} (see Supplemental Materials) to establish
that 
the prefactor of $F_\mrm{sc}^2$
of Eq.~(\ref{eq:7}) enters into $-\tensor{P}(0)$ with the opposite sign compared to the diamagnetic current.
Once we know the form for $\tensor{P}(0)$ we can make an ansatz
for the form of $\tensor{P}(Q)$ (compatible
with BCS theory when $\Delta_\mrm{pg} \equiv 0$).
While there is no unique inference for $\tensor{P}(Q)$ away from $Q=0$, 
we depend on
the explicit satisfaction
of the transverse and f-sum rules to support our ansatz. 
Our precise form for $\tensor{P}(0)$ and
our ansatz for $\tensor{P}(Q)$ are given by   

\begin{widetext}
\begin{equation}
-\tensor{P}(0) = 
-2\sum_KG^2_K\frac{\partial\xi_\k}{\partial \k}
\frac{\partial\xi_\k}{\partial \k}
+
2\sum_KF_{\mrm{pg},K}^2 \frac{\partial\xi_\k^\mrm{pg}}{\partial \k}
\frac{\partial\xi_\k}{\partial \k}
-
2\sum_KF_{\mrm{sc},K}^2 \left[
\frac{\partial\xi_\k}{\partial \k}
\frac{\partial\xi_\k}{\partial \k}
-
\Lambda^\mrm{sc} \frac{\partial\xi_\k^\mrm{pg}}{\partial \k}
\frac{\partial\xi_\k}{\partial \k} \Delta_\mrm{pg}^2 (G_0^\mrm{pg}(K))^2
\right],
\label{eq:8}
\end{equation}

\begin{eqnarray}
\tensor{P}(Q) &=& 2\sum_K\frac{\partial\xi_
{\textbf{k}+\textbf{q}/2}}{\partial\textbf{k}}
\frac{
\partial\xi_{\textbf{k}+\textbf{q}/2}}{\partial\textbf{k}}G_KG_{K+Q} 
- 
2\sum_K\frac{\partial\xi^\mrm{pg}_{\textbf{k}+\textbf{q}/2}}{\partial\textbf{k}}
\frac{\partial\xi_{\textbf{k}+\textbf{q}/2}}{\partial\textbf{k}}F_{\mrm{pg},K}
F_{\mrm{pg},K+Q}
\nonumber \\
&+&
2 \sum_KF_{\mrm{sc},K}F_{\mrm{sc},K+Q} \left[ 
\frac{\partial\xi_{\textbf{k}+\textbf{q}/2}}{\partial\textbf{k}}
\frac{\partial\xi_{\textbf{k}+\textbf{q}/2}}
{\partial\textbf{k}}
-
\Lambda^\mrm{sc} \frac{\partial\xi_{\textbf{k}+\textbf{q}/2}^\mrm{pg}}
{\partial\textbf{k}}
\frac{\partial\xi_{\textbf{k}+\textbf{q}/2}}
{\partial\textbf{k}}
\Delta_\mrm{pg}^2 G_0^\mrm{pg}(K)G_0^\mrm{pg}(K+Q) \right],
\label{eq:9b}
\end{eqnarray}
\end{widetext}
where we only consider the transverse response $P_t$ below $T_c$ (the
longitudinal part of $\tensor{P}(Q)$ is correct in the normal phase,
but requires collective mode corrections for $T < T_c$).

The quantities 
$\tensor{P}(0)$
and
$\big(\frac{\tensor{n}}{m}\big)_\mrm{dia}$, 
are, however, all that is needed to
deduce an expression for the superfluid density 
$\frac{n_s}{m} \equiv \big(\frac{n}{m}\big)_\mrm{dia} - P_t(0)$ 
in both the arcs and pockets model,
\begin{equation}
\frac{n_s}{m} =
4\sum_KF_{\mrm{sc},K}^2 \big[
\frac{\partial\xi_\k}{\partial \k}
\frac{\partial\xi_\k}{\partial \k}
-
\Lambda^\mrm{sc} \frac{\partial\xi_\k^\mrm{pg}}{\partial \k}
\frac{\partial\xi_\k}{\partial \k} \Delta_\mrm{pg}^2 (G_0^\mrm{pg}(K))^2
\big].
\label{eq:9}
\end{equation}
It is interesting to note that in the review on YRZ \cite{RiceReview}, a concern was raised that
the penetration depth (or $n_s/m$) which appears in the YRZ literature is missing a vertex correction.
Here, with Eqs.~(\ref{eq:7})-(\ref{eq:8}),
we have established the form for such a vertex correction.\footnote{It 
should be noted that this vertex correction (which, for the
pockets scenario, depends on the cross term $\Delta_\mrm{sc}^2 
\Delta_\mrm{pg}^2$)
introduces an effective gap shape which differs from the simple
$d$-wave form.}

\begin{figure*}
\includegraphics[width=6.0in,clip]
{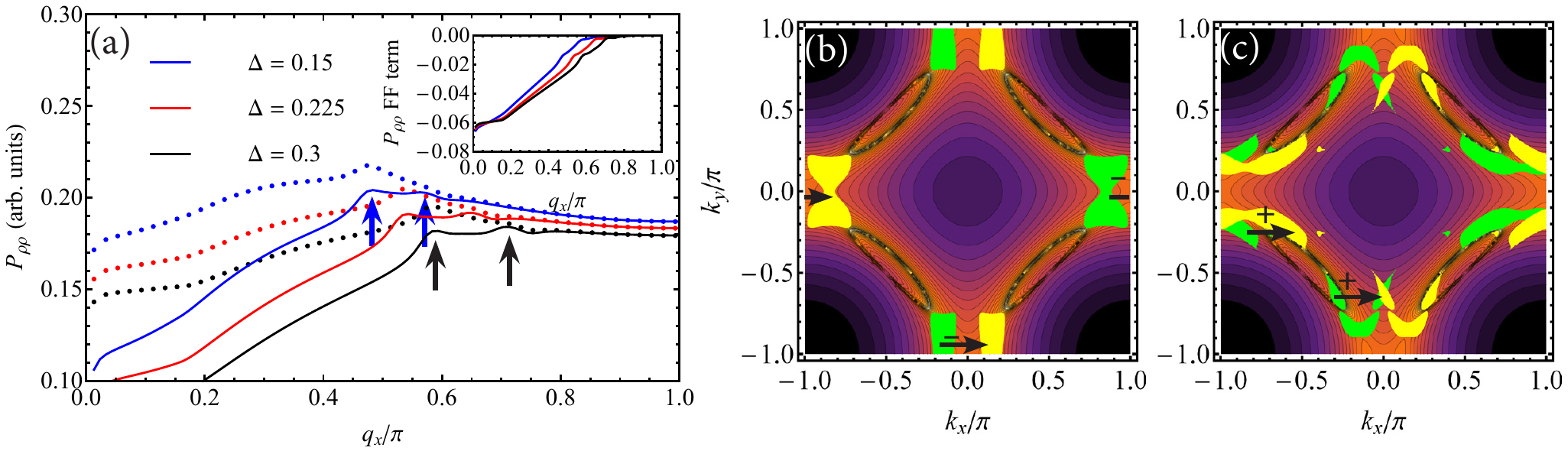}
\caption{(a) Normal state $\omega = 0$, $q_y = 0$ charge susceptibility
$P_{\rho\rho}$ with
(solid) and without (dotted) vertex corrections. The arrows indicate that
a second peak is present in the former case. Here we follow the band
structure used in Ref.~\onlinecite{Comin}, and use
$T = 0.01$ and broadening $\eta = 0.01$\cite{Comin} 
to study a low-temperature system.  The doping $p = 0.12$, and chemical
potential $\mu$ is fixed by the Luttinger sum rule \cite{YRZ}.
These values are normalized to $t$, the primary 
single-particle dispersion parameter \cite{YRZ,Comin}. The inset shows
the contribution of the vertex term ($F_\mrm{pg}F_\mrm{pg}$) to
$P_{\rho\rho}$.
(b-c) Plots of the momentum phase space contributions to
$P_{\rho\rho}(\omega=0,\q = (0.3 \pi,0))$ for $\Delta = 0.15$,
overlaid on contour plots of the spectral function $A_\mrm{YRZ}(\omega =
0,\k)$. Shown are green $\k$ (``origin'') regions and yellow $\k + \q$
(''destination'') regions for which the integrand magnitude 
is greater than a set threshold. (b) shows 
contributions from the vertex term greater than a threshold of $0.008$, while
(c) shows single fermionic bubble ($GG$) contributions greater than a threshold
of $0.02$. The plus and minus signs in (b) and (c) indicate the sign of the
phase space contribution. } 
\label{fig:1}
\end{figure*}

In the normal state
and for both the pockets and arcs model, one can show that
the density-density response function is given by
\begin{equation}
P_{\rho\rho}(Q) = {2}\sum_K\Bigg(
\Big[G_KG_{K+Q} \Big]
+
\Big[F_{\mrm{pg},K}F_{\mrm{pg},K+Q}\Big]\Bigg).\label{eq:fullP}
\end{equation}
This equation will be used in the remainder of this paper
to establish the way in which previously omitted\cite{Comin} 
vertex corrections in the second term impact
the charge response functions. 
We restrict calculations to $T > T_c$ so as to avoid
complications from collective
modes in the presence of pseudogap effects.
We can similarly address the quasi-particle interference 
pattern\cite{ourqpi,ourqpi2} of STM,
as well as the complex conductivity and diamagnetic 
susceptibility,\cite{ouroptical,ourthz,Ourconduct,ourdia}
all
of which
are given in the Supplementary Materials.

Finally, the spin current and density response functions
can be similarly deduced.
Indeed $\big(\tensor{\frac{n}{m}}\big)_\mrm{dia}$
in Eq.~(\ref{eq:7}), appears in the constraining
sum rules on the vertex corrections. The spin-current correlation function is
given by $\tensor P(Q)$ with a sign change in front of $F_\mrm{sc}$, reflecting
the absence of a Meissner effect, as the spin pairing is
assumed to be singlet. Above $T_c$, the bare dynamic susceptibility
$\chi^\mrm{spin}(Q)$ 
is the same as the expression in
Eq.~(\ref{eq:fullP}), where the second term represents the vertex corrections.
These are necessary to ensure that the formation of singlets leads to a 
normal state gap
in the spin excitation spectrum, which is not fully accounted for 
by the first term.
Below $T_c$, $\chi^\mrm{spin}(Q)$ must include the vertex
corrections associated with $\Lambda^\mrm{sc}$, but there are no collective mode effects.

\textit{Consistency with sum rules.}
The normal state f-sum rule (on a lattice) provides a strong
constraint on the charge susceptibility in Eq.~(\ref{eq:fullP})
of the form:
\begin{equation}
\int \frac{d\w}{\pi}\left(-\w {\rm Im} P_{\rho\rho}(Q)\right)
=2\sum_{\k} (\xi_{\k+\q}+\xi_{\k-\q}-2\xi_\k) n_\k,
\label{eq:14} 
\end{equation}
with $n_\k=\mean{\hat c^\dag_{\k,\sigma} \hat c_{\k,\sigma}}$ (here $Q=(\w+i0^+,\q)$).
In the YRZ pockets model, the left-hand side of Eq.~\eqref{eq:14} gives
\begin{equation}
\begin{split}
\int \frac{d\w}{\pi}&\left(-\w {\rm Im}P_{\rho\rho}(Q)\right)=
2\sum_{\k,\alpha=\pm,i=1,2}(-1)^{i-1} f(E_{i,\alpha}) \\ &\times \frac{(E_{i,\alpha}+\xi^\mrm{pg}_\alpha)(E_{i,\bar\alpha}+E_{\bar i,\bar\alpha}+\xi^\mrm{pg}_{\bar\alpha}-E_{i,\alpha})+\Delta_\mrm{pg}^2}{E_{1,\alpha}-E_{2,\alpha}},
\label{eq:fsum} 
\end{split}
\end{equation}
where 
$E_{i,\k}=\frac{1}{2}\left(\xi_\k-\xi^\mrm{pg}_\k+(-1)^{i-1}
\sqrt{(\xi_\k+\xi^\mrm{pg}_\k)^2+4\Delta_\mrm{pg}^2}\right)$, $i=1,2$ are
the poles of the YRZ Green's function, and we define $\alpha=\pm$ to represent
$\k\pm \q/2$. We introduce
$\bar\alpha=-\alpha$ and $\bar 1=2$, $\bar 2=1$.
Using the two identities
$E_{i,\bar\alpha}+E_{\bar i,\bar\alpha}+\xi^\mrm{pg}_{\bar\alpha}=\xi_{\bar
\alpha}$,
and
$E_{i,\alpha}(E_{i,\alpha}+\xi^\mrm{pg}_\alpha)=\xi_\alpha(E_{i,\alpha}+\xi^
\mrm{pg}_\alpha)+\Delta_\mrm{pg}^2$,
as well as the change of variable $\k\to \k-\alpha\q/2$, we find the right-hand side
of Eq.~(\ref{eq:fsum}) reads
\begin{equation}
\begin{split}
2\sum_{\k,i=1,2}& (-1)^{i-1} f(E_{i,\k})\frac{(E_{i,\k}+\xi^\mrm{pg}_\k)(\xi_{\k+\q}+\xi_{\k-\q}-2\xi_\k)}{E_{1,\k}-E_{2,\k}}\\
&=2\sum_{\k} (\xi_{\k+\q}+\xi_{\k-\q}-2\xi_\k) n_\k,\nonumber
\end{split}
\end{equation}
which is the longitudinal f-sum rule for YRZ in the normal state, since 
\begin{equation}
n_\k=\frac{(E_{1,\k}+\xi^\mrm{pg}_\k)f(E_{1,\k})-(E_{2,\k}+\xi^\mrm{pg}_\k)f(E_{2,\k})}{E_{1,\k}-E_{2,\k}}.\nonumber
\end{equation}
The derivation in the arc case is essentially the same, with  $\xi^\mrm{pg}_\k=\xi_\k$ and $E_{\k,2}=-E_{\k,1}$.

It should also be clear
that the f-sum rule in Eq.~(\ref{eq:14}) assumes a more subtle
form in the presence of a lattice, as it does not directly
depend on 
$\big(\frac{n}{m} \big)_\mrm{dia} \times q^2$. One should think of
$\big(\frac{n}{m} \big)_\mrm{dia}$ as reflecting
a $q \rightarrow 0$
limit of the response functions, 
whereas the f-sum rule was proved  
above to be
valid for all $q$.\footnote{Despite this rather strong 
validation of the generalized response functions, we note
that for both the arcs and pockets
models one will not satisfy the compressibility sum rule. This can be
traced to the assumed, simplified form for the self energies
which are not functions of $\w+\mu$ as they would be expected to be
from the gauge invariance of the microscopic Hamiltonian.}
Finally, the transverse sum rule
is shown in the Supplementary Materials to be consistent with Eq.~\eqref{eq:9b}.

\vskip2mm
\textit{Numerical results and discussion.}
We turn now to a quantification of vertex corrections 
and show that this  
leads to a much better
understanding of their physical nature and origin. 
Results using the method of calculation presented in 
Ref.~\onlinecite{Comin}
arising from only including the so-called ``bubble" contribution
are shown as dotted lines in the left panel
of Figure \ref{fig:1} with a single peak. 
They are compared with the full charge susceptibility in
Eq.~(\ref{eq:fullP}), shown as solid lines.
As the gap $\Delta_\mrm{pg}$ increases, the magnitude of the (negative) vertex correction term tends to increase,
as indicated in the inset. Importantly, the arrows indicate that this 
introduces a second peak which is of equal magnitude for larger $\Delta_\mrm{pg}$. 
In the small $\bf{q}$ regime vertex corrections remove almost half the 
weight found in the dotted line bubble contribution.

To understand the physical cause of these vertex corrections, we plot in the middle
panel a color contour figure of
the dominant phase space contributions to the integrand in Eq.~(\ref{eq:fullP}) deriving
from the vertex corrections for a fixed $\mathbf{q}$,
as indicated by the arrows.
These corrections are rather strongly localized to the antinodes. To elucidate
this
we note that $F_\mrm{pg}$ given in Eq.~(\ref{eq:3})
can be interpreted as a bosonic Green's function since its spectral function exhibits the
appropriate sign change when $\omega \rightarrow - \omega$. This bosonic degree of
freedom is naturally associated with fermionic pairing and is expected, then, to
reside near the antinodes and to increase in
magnitude as pairing gets stronger. We may then interpret the vertex corrections in
Eq.~(\ref{eq:fullP}) as arising from the spin singlets 
in a resonating valence bond (RVB)\cite{RVB,Andersonbook}
context,
leading to a picture which is
not so different from that expounded in Ref.~\onlinecite{Larkin}.
By contrast, the right hand panel indicates the phase space contributions arising from
the simple $GG$ ``bubble" which tend not to be so relatively strong near the antinode.

\textit{Conclusions.} 
All of the results presented here follow rather directly from the form
of the self energy $\Sigma_\mrm{pg}$ which, through a Ward identity, will
affect correlation functions in a way which we have just interpreted.
An emerging theme is that even though there has been no explicit
reference to the spin singlets of RVB, these arguments indicate that
one has a two-constituent system. Ignoring vertex corrections in the
case of the charge susceptibility is largely ignoring this bosonic 
constituent. Indeed, even in thermodynamics, not just in the spin
and charge response functions, one should expect some
residue of bosonic degrees of freedom both directly and indirectly through
the gap which they present to the fermionic sector.

Analogous studies are presented  for the arcs
scenario, except that there are no ``hot spots" or pocket tips
to lead to sharp peaks in the charge susceptibility.\cite{Yanspaper} This is
illustrated in the Supplementary Materials.
But more significant is the similarity which
allows a consolidation of two (at first sight) rather different
approaches to the cuprate pseudogap: the pockets model
of YRZ\cite{RVB,YRZ} and the arcs model of BCS-BEC crossover.\cite{ourreview} Both of these
have two distinct gaps corresponding to the condensed and
non-condensed pairs, although
the YRZ is more specific by associating singlet 
pairing with antiferromagnetic correlations.
As noted in Ref.~\onlinecite{YRZ}, throughout
the temperature range, ``both gaps keep their
own identity". 
For this reason, among others,
these two-gap approaches are 
distinguished from phase fluctuation scenarios,\cite{RiceMag} and allow the
general and consistent treatment of response functions presented here.

\vskip2mm
This work is supported by NSF-MRSEC Grant
0820054.


\bibliography{Review2}

\clearpage




\section*{Supplementary material}

# Supplemental materials

## VERTEX CORRECTIONS IN CONDUCTIVITY, DIAMAGNETISM AND QUASI-PARTICLE INTERFERENCE

Once one has the current-current response function with vertex corrections

$$\overset{\leftrightarrow}{P}(Q) = 2\sum_K \frac{\partial \xi_{\mathbf{k}+\mathbf{q}/2}}{\partial \mathbf{k}} \frac{\partial \xi_{\mathbf{k}+\mathbf{q}/2}}{\partial \mathbf{k}} G_K G_{K+Q} - 2\sum_K \frac{\partial \xi^{\text{pg}}_{\mathbf{k}+\mathbf{q}/2}}{\partial \mathbf{k}} \frac{\partial \xi_{\mathbf{k}+\mathbf{q}/2}}{\partial \mathbf{k}} F_{\text{pg},K} F_{\text{pg},K+Q}$$
$$+ 2\sum_K F_{\text{sc},K} F_{\text{sc},K+Q} \left[ \frac{\partial \xi_{\mathbf{k}+\mathbf{q}/2}}{\partial \mathbf{k}} \frac{\partial \xi_{\mathbf{k}+\mathbf{q}/2}}{\partial \mathbf{k}} - \Lambda^{\text{sc}} \frac{\partial \xi^{\text{pg}}_{\mathbf{k}+\mathbf{q}/2}}{\partial \mathbf{k}} \frac{\partial \xi_{\mathbf{k}+\mathbf{q}/2}}{\partial \mathbf{k}} \Delta^2_{\text{pg}} G^{\text{pg}}_0(K) G^{\text{pg}}_0(K+Q) \right], \quad (1)$$

the complex conductivity and diamagnetic susceptibility are given by

$$\sigma_{xx}(\omega) = -\lim_{\mathbf{q}\to 0} \frac{P_{xx}(\mathbf{q},\omega) + (n/m)^{\text{dia}}_{xx}}{i\omega} \qquad \chi^{\text{dia}} = -\lim_{q_y\to 0} \text{Re}\left[\frac{P_{xx}(\mathbf{q},\omega=0) + (n/m)^{\text{dia}}_{xx}}{\mathbf{q}^2}\right]_{q_x=q_z=0}. \quad (2)$$

The quasi-particle interference pattern reflects the Fourier transform of the changes in the local density of states (LDOS) due to the presence of impurities, which are characterized by a potential $U_0$. Here, too, because of the pg self-energy, vertex contributions enter. The LDOS with vertex corrections is given by

$$\delta n(\omega, \mathbf{q}) \propto \text{Im} \int \frac{d^2 k}{(2\pi)^2} \Big( G_K G_{K+Q_0} - F^{\text{sc}}_K F^{\text{sc}}_{K+Q_0} \times [1 - \Lambda^{\text{sc}} G^{\text{pg}}_0(K) G^{\text{pg}}_0(K+Q_0) \Delta^2_{\text{pg}}] - F^{\text{pg}}_K F^{\text{pg}}_{K+Q_0} \Big),$$

with $Q_0 \equiv (0, \mathbf{q})$ and $K = (\omega, \mathbf{k})$.

## WARD IDENTITIES AND VERTEX FUNCTIONS

### Normal Pseudogap phase

The Ward identity for the full vertex function $\Gamma_\mu(K,Q) = (\Gamma_0(K,Q), \boldsymbol{\Gamma}(K,Q))$ in the normal phase is given by

$$-\Omega \Gamma_0(K,Q) - i\,\text{div}_{\mathbf{q}}.\boldsymbol{\Gamma}(K,Q) = G^{-1}_K - G^{-1}_{K+Q},$$
$$= -\Omega + \xi_{\mathbf{k}+\mathbf{q}} - \xi_{\mathbf{k}} + \Delta^2_{\text{pg}} \left( \frac{1}{\omega + \Omega + \xi^{\text{pg}}_{\mathbf{k}+\mathbf{q}}} - \frac{1}{\omega + \xi^{\text{pg}}_{\mathbf{k}}} \right), \quad (3)$$

where $Q = (\Omega, \mathbf{q})$, $K = (\omega, \mathbf{k})$, and $\text{div}_{\mathbf{q}}.\boldsymbol{\Gamma}(K,Q)$ represent the Fourier transform of the divergence of $\boldsymbol{\Gamma}$, generalized to the lattice. (Note that $\text{div}_{\mathbf{q}}.\boldsymbol{\Gamma}(K,Q) \neq i\mathbf{q}.\boldsymbol{\Gamma}(K,Q)$ for lattice models, unless $q \to 0$.) The first line of Eq (3) is exact due to charge conservation, whereas the second line follows from the choice of the self-energy (see main text). For the bare vertex function $\gamma_\mu(K,Q) = (\gamma_0(K,Q), \boldsymbol{\gamma}(K,Q))$, this corresponds to

$$\gamma_0(K,Q) = 1,$$
$$-i\,\text{div}_{\mathbf{q}}.\boldsymbol{\gamma}(K,Q) = \xi_{\mathbf{k}+\mathbf{q}} - \xi_{\mathbf{k}}. \quad (4)$$

From Eq. (3), we can infer that the vertex function is given by

$$\Gamma_0(K,Q) = \gamma_0(K,Q) \left[ 1 + \frac{\Delta^2_{\text{pg}}}{(\omega + \Omega + \xi^{\text{pg}}_{\mathbf{k}+\mathbf{q}})(\omega + \xi^{\text{pg}}_{\mathbf{k}})} \right],$$
$$\boldsymbol{\Gamma}(K,Q) = \boldsymbol{\gamma}(K,Q) - \boldsymbol{\gamma}^{\text{pg}}(K,Q) \frac{\Delta^2_{\text{pg}}}{(\omega + \Omega + \xi^{\text{pg}}_{\mathbf{k}+\mathbf{q}})(\omega + \xi^{\text{pg}}_{\mathbf{k}})}, \quad (5)$$

where the 'pg vertex function' $\boldsymbol{\gamma}^{\mathrm{pg}}(K,Q)$ is chosen such that

$$-i\,\mathrm{div}_{\mathbf{q}}.\boldsymbol{\gamma}^{\mathrm{pg}}(K,Q) = \xi^{\mathrm{pg}}_{\mathbf{k+q}} - \xi^{\mathrm{pg}}_{\mathbf{k}}. \tag{6}$$

The charge-charge and current-current correlation functions are then given by

$$\begin{aligned}P_{\rho\rho}(Q) &= 2\sum_K \Gamma_0(K,Q)\gamma_0(K+Q,-Q)G_K G_{K+Q}, \\ \overset{\leftrightarrow}{P}(Q) &= 2\sum_K \boldsymbol{\Gamma}(K,Q)\boldsymbol{\gamma}(K+Q,-Q)G_K G_{K+Q},\end{aligned} \tag{7}$$

corresponding to the normal phase version of Eqs. (7) and (11) of the main text. The choice of the vertex function in Eq. (5) is non-trivial, but this inference can be checked against both conservation of particles (through the Ward identity, which provides the vertex construction), and less trivially by the different sum-rules.

### Superconducting phase

Following standard textbooks, the Ward identity in the superconducting phase is given, using Nambu spinor notation

$$-\Omega\,\Gamma_0(K,Q) - i\,\mathrm{div}_{\mathbf{q}}.\boldsymbol{\Gamma}(K,Q) = \tau_3 \mathcal{G}^{-1}(K) - \mathcal{G}^{-1}(K+Q)\tau_3, \tag{8}$$

where $\tau_3$ is the third Pauli matrix (we will also use $\tau_0$ for the identity in Nambu space), and

$$\mathcal{G}(K) = \begin{pmatrix} G(K) & F_{\mathrm{sc}}(K) \\ F_{\mathrm{sc}}(K) & -G(-K) \end{pmatrix}, \tag{9}$$

is the Nambu matrix Green's function. Note that $\Gamma_\mu$ is now a matrix. For the YRZ model, Eq. (8) becomes

$$-\Omega\,\Gamma_0(K,Q) - i\,\mathrm{div}_{\mathbf{q}}.\boldsymbol{\Gamma}(K,Q) = -\Omega\,\tau_3 + \tau_0\,(\xi_{\mathbf{k+q}} - \xi_{\mathbf{k}}) + \begin{pmatrix} \Delta_{\mathrm{pg}}^2 \left( \frac{1}{\omega+\Omega+\xi^{\mathrm{pg}}_{\mathbf{k+q}}} - \frac{1}{\omega+\xi^{\mathrm{pg}}_{\mathbf{k}}} \right) & 2\Delta_{\mathrm{sc}} \\ -2\Delta_{\mathrm{sc}} & \Delta_{\mathrm{pg}}^2 \left( \frac{1}{-\omega-\Omega+\xi^{\mathrm{pg}}_{\mathbf{k+q}}} - \frac{1}{-\omega+\xi^{\mathrm{pg}}_{\mathbf{k}}} \right) \end{pmatrix}. \tag{10}$$

We are interested in the transverse current-current correlation function, which is insensitive to the collective mode physics. This allows us to discard the $\Delta_{\mathrm{sc}}$ part of the Ward identity (which corresponds to the collective mode singularity of the longitudinal vertex function). We thus infer the transverse vertex function

$$\boldsymbol{\Gamma}_t(K,Q) = \tau_0 \boldsymbol{\gamma}_t(K,Q) - \boldsymbol{\gamma}_t^{\mathrm{pg}}(K,Q) \begin{pmatrix} \frac{\Delta_{\mathrm{pg}}^2}{(\omega+\Omega+\xi^{\mathrm{pg}}_{\mathbf{k+q}})(\omega+\xi^{\mathrm{pg}}_{\mathbf{k}})} & 0 \\ 0 & \frac{\Delta_{\mathrm{pg}}^2}{(-\omega-\Omega+\xi^{\mathrm{pg}}_{\mathbf{k+q}})(-\omega+\xi^{\mathrm{pg}}_{\mathbf{k}})} \end{pmatrix}. \tag{11}$$

The current-current correlation function in the superconducting phase is given by

$$\overset{\leftrightarrow}{P}(Q) = \sum_K \mathrm{Tr}\Big\{ \boldsymbol{\Gamma}(K,Q)\mathcal{G}(K)\boldsymbol{\gamma}(K+Q,-Q)\mathcal{G}(K+Q) \Big\}, \tag{12}$$

where the trace is over the Nambu indices, which gives the transverse current-current correlation function quoted in the main text (Eq.7). We will show next that this choice for $\overset{\leftrightarrow}{P}(Q)$ is consistent with the transverse sum rule.

### TRANSVERSE SUM RULE

The transverse sum-rule is given by $\frac{n_n}{m} = \lim_{\mathbf{q}\to 0} \int \frac{d\omega}{\pi} \left( -\frac{\mathrm{Im}P_t(i\Omega\to\omega+i0^+,\mathbf{q})}{\omega} \right)$ (with $P_t(Q)$ the transverse part of $\overset{\leftrightarrow}{P}(Q)$), where $\overset{\leftrightarrow}{P}(Q)$ is computed with Matsubara frequencies $i\Omega$. Given the definition of the superfluid density, this sum rule imposes the requirement that $\overset{\leftrightarrow}{P}(Q)$ discussed above obeys the following Kramers-Kronig relation

$$\lim_{\mathbf{q}\to 0} \int \frac{d\omega}{\pi} \left( -\frac{\mathrm{Im}P_t(\omega,\mathbf{q})}{\omega} \right) = -P_t(Q=0). \tag{13}$$

To demonstrate this Kramers-Kronig consistency, we will use the fact that we can rewrite all the Green's functions in $P_t(Q)$ and $P_t(0)$ using spectral functions, and that term by term they give identical contributions.

For example, $G(K)$ can be written as (with $i\omega$ a fermionic Matsubara frequency)

$$G(K) = \sum_i \frac{R_{\mathbf{k}}^i}{i\omega - E_{\mathbf{k}}^i}, \tag{14}$$

where $E_{\mathbf{k}}^i$ are the poles of $G(K)$ and $R_{\mathbf{k}}^i$ the corresponding residues (there are four of them for YRZ). For the other combinations of $G$, $G_0$ and $F$'s, the number of poles is the same, the only difference is in the residues. To simplify the notations, we can rewrite

$$P_t(Q) = \sum_K \sum_a X_a(\mathbf{k}+\mathbf{q}/2) g_a(K) g_a(K+Q), \tag{15}$$

where $a = \{1,2,3,4\}$ indicates the four contributions to $P_t$, and $X_1(\mathbf{k}+\mathbf{q}/2) = 2\frac{\partial \xi_{\mathbf{k}+\mathbf{q}/2}}{\partial \mathbf{k}}\frac{\partial \xi_{\mathbf{k}+\mathbf{q}/2}}{\partial \mathbf{k}}$, and $X_2(\mathbf{k}+\mathbf{q}/2) = -2\frac{\partial \xi_{\mathbf{k}+\mathbf{q}/2}^{\mathrm{pg}}}{\partial \mathbf{k}}\frac{\partial \xi_{\mathbf{k}+\mathbf{q}/2}}{\partial \mathbf{k}}$ and $X_3(\mathbf{k}+\mathbf{q}/2) = 2\frac{\partial \xi_{\mathbf{k}+\mathbf{q}/2}}{\partial \mathbf{k}}\frac{\partial \xi_{\mathbf{k}+\mathbf{q}/2}}{\partial \mathbf{k}}$ and $X_4(\mathbf{k}+\mathbf{q}/2) = -2\Lambda^{\mathrm{sc}}\frac{\partial \xi_{\mathbf{k}+\mathbf{q}/2}^{\mathrm{pg}}}{\partial \mathbf{k}}\frac{\partial \xi_{\mathbf{k}+\mathbf{q}/2}}{\partial \mathbf{k}}$. Here $g_1(K) = G_K$, $g_2(K) = F_{\mathrm{pg},K}$, $g_3(K) = F_{\mathrm{sc},K}$, $g_4(K) = \Delta_{\mathrm{pg}}^2 G_0^{\mathrm{pg}}(K) F_{\mathrm{sc},K}$.

We will first compute the left-hand side of Eq. (13) and then show that the right-hand side is consistent with this result.

**Calculation of** $\lim_{\mathbf{q}\to 0}\int \frac{d\omega}{\pi}\left(-\frac{\operatorname{Im}P_t(\omega,\mathbf{q})}{\omega}\right)$

It is useful to introduce a relation involving the relevant Matsubara summations

$$h_a(i\Omega) = T\sum_{i\omega} g_a(K) g_a(K+Q) = T\sum_{i\omega}\int_{x,y}\frac{A_{a,\mathbf{k}}(x)}{i\omega - x}\frac{A_{a,\mathbf{k}+\mathbf{q}}(y)}{i\omega + i\Omega - y} = \int_{x,y}\frac{A_{a,\mathbf{k}}(x)A_{a,\mathbf{k}+\mathbf{q}}(y)}{i\Omega + x - y}(f(x) - f(y)), \tag{16}$$

where $f(x)$ is the Fermi distribution and $A_{a,\mathbf{k}}(x)$ is the spectral function of $g_a(K)$, i.e.,

$$g_a(K) = \int \frac{dx}{2\pi}\frac{A_{a,\mathbf{k}}(x)}{i\omega - x}. \tag{17}$$

To prove the sum rule, we need the calculation of

$$\int \frac{d\omega}{\pi}\left(-\frac{\operatorname{Im}h_a(\omega+i0^+)}{\omega}\right) = \int \frac{dx}{2\pi}\frac{dy}{2\pi}A_{a,\mathbf{k}}(x)A_{a,\mathbf{k}+\mathbf{q}}(y)\frac{f(x) - f(y)}{y - x} = H_a(\mathbf{k},\mathbf{q}). \tag{18}$$

as well as the limit $\mathbf{q}\to 0$ of $\sum_{\mathbf{k}}X_a(\mathbf{k}+\mathbf{q}/2)H_a(\mathbf{k},\mathbf{q})$. First, note that the limit $\lim_{\mathbf{q}\to 0}X_a(\mathbf{k}+\mathbf{q}/2) = X_a(\mathbf{k})$ is well defined. However, one must exercise care with this limit for the product of the two spectral function $A_{a,\mathbf{k}}(x)A_{a,\mathbf{k}+\mathbf{q}}(y)$. The spectral functions are explicitly given by

$$A_{a,\mathbf{k}}(x) = 2\pi\sum_i R_{\mathbf{k}}^{a,i}\delta(x - E_{\mathbf{k}}^{a,i}), \tag{19}$$

and we thus have

$$H_a(\mathbf{k},\mathbf{q}) = \sum_{i,j} R_{\mathbf{k}}^{a,i} R_{\mathbf{k}+\mathbf{q}}^{a,j}\frac{f(E_{\mathbf{k}}^{a,i}) - f(E_{\mathbf{k}+\mathbf{q}}^{a,j})}{E_{\mathbf{k}+\mathbf{q}}^{a,j} - E_{\mathbf{k}}^{a,i}}. \tag{20}$$

The limit $\mathbf{q}\to 0$ of $R_{\mathbf{k}+\mathbf{q}}^{a,j}$ is well defined, as is that of $\frac{f(E_{\mathbf{k}}^{a,i}) - f(E_{\mathbf{k}+\mathbf{q}}^{a,j})}{E_{\mathbf{k}+\mathbf{q}}^{a,j} - E_{\mathbf{k}}^{a,i}}$ for $i\neq j$. Thus

$$\lim_{\mathbf{q}\to 0} H_a(\mathbf{k},\mathbf{q}) = \sum_{i\neq j} R_{\mathbf{k}}^{a,i} R_{\mathbf{k}}^{a,j}\frac{f(E_{\mathbf{k}}^{a,i}) - f(E_{\mathbf{k}}^{a,j})}{E_{\mathbf{k}}^{a,j} - E_{\mathbf{k}}^{a,i}} - \sum_i (R_{\mathbf{k}}^{a,i})^2 f'(E_{\mathbf{k}}^{a,i}). \tag{21}$$

The first part of the sum-rule thus gives

$$\lim_{\mathbf{q}\to 0}\int \frac{d\omega}{\pi}\left(-\frac{\operatorname{Im}P_t(\omega,\mathbf{q})}{\omega}\right) = \sum_a \sum_{\mathbf{k}} X_a(\mathbf{k})\left(\sum_{i\neq j} R_{\mathbf{k}}^{a,i} R_{\mathbf{k}}^{a,j}\frac{f(E_{\mathbf{k}}^{a,i}) - f(E_{\mathbf{k}}^{a,j})}{E_{\mathbf{k}}^{a,j} - E_{\mathbf{k}}^{a,i}} - \sum_i (R_{\mathbf{k}}^{a,i})^2 f'(E_{\mathbf{k}}^{a,i})\right). \tag{22}$$

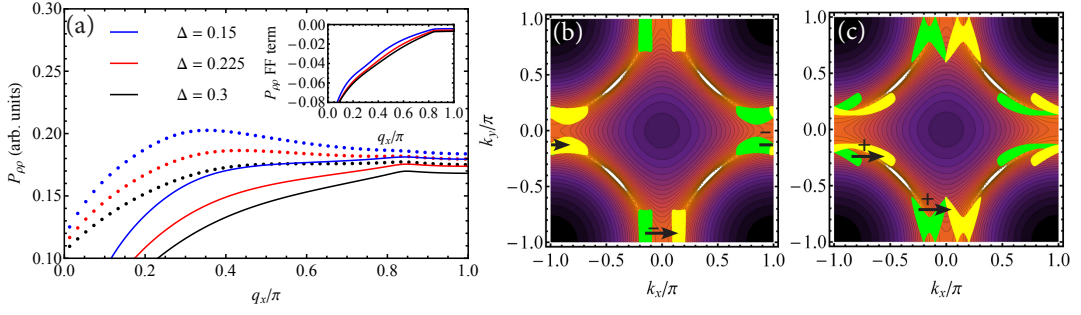

Figure 1: Arc Scenario Results–(a) Normal state $\omega = 0$ charge susceptibility with (solid) and without (dotted) vertex corrections. (b-c) These panels indicate the momentum phase space contributions broken down into (b) the vertex term ($F_{\text{pg}}F_{\text{pg}}$) and (c) the single fermionic bubble ($GG$). All parameters are the same as in Figure 1 of the main paper.

**Calculation of $P_t(0)$**

We start from

$$P_t(0) = \sum_a \sum_K X_a(\mathbf{k}) g_a(K)^2, \tag{23}$$

and we first rewrite the Matsubara sum using the spectral representation

$$T \sum_{i\omega} g_a(K)^2 = T \sum_{i\omega} \int \frac{dx}{2\pi} \frac{dy}{2\pi} \frac{A_{a,\mathbf{k}}(x)}{i\omega - x} \frac{A_{a,\mathbf{k}}(y)}{i\omega - y}. \tag{24}$$

We next integrate over $x$ and $y$, considering that the Matsubara sum might contain double poles. A straightforward calculation gives

$$\begin{aligned}
T \sum_{i\omega} g_a(K)^2 &= T \sum_{i\omega} \sum_{i,j} \int \frac{dx}{2\pi} \frac{dy}{2\pi} \frac{R_{\mathbf{k}}^{a,i} \delta(x - E_{\mathbf{k}}^{a,i})}{i\omega - x} \frac{R_{\mathbf{k}}^{a,j} \delta(y - E_{\mathbf{k}}^{a,j})}{i\omega - y}, \\
&= T \sum_{i\omega} \left( \sum_{i \neq j} \frac{R_{\mathbf{k}}^{a,i} R_{\mathbf{k}}^{a,j}}{(i\omega - E_{\mathbf{k}}^{a,i})(i\omega - E_{\mathbf{k}}^{a,j})} + \sum_i \frac{(R_{\mathbf{k}}^{a,i})^2}{(i\omega - E_{\mathbf{k}}^{a,i})^2} \right), \\
&= \sum_{i \neq j} R_{\mathbf{k}}^{a,i} R_{\mathbf{k}}^{a,j} \frac{f(E_{\mathbf{k}}^{a,i}) - f(E_{\mathbf{k}}^{a,j})}{E_{\mathbf{k}}^{a,i} - E_{\mathbf{k}}^{a,j}} + \sum_i (R_{\mathbf{k}}^{a,i})^2 f'(E_{\mathbf{k}}^{a,i}).
\end{aligned} \tag{25}$$

Therefore, we find

$$\begin{aligned}
P_t(0) &= \sum_a \sum_{\mathbf{k}} X_a(\mathbf{k}) \left( \sum_{i \neq j} R_{\mathbf{k}}^{a,i} R_{\mathbf{k}}^{a,j} \frac{f(E_{\mathbf{k}}^{a,i}) - f(E_{\mathbf{k}}^{a,j})}{E_{\mathbf{k}}^{a,i} - E_{\mathbf{k}}^{a,j}} + \sum_i (R_{\mathbf{k}}^{a,i})^2 f'(E_{\mathbf{k}}^{a,i}) \right), \\
&= -\lim_{\mathbf{q} \to 0} \int \frac{d\omega}{\pi} \left( -\frac{\text{Im} P_t(\omega, \mathbf{q})}{\omega} \right),
\end{aligned} \tag{26}$$

which shows consistency with the sum rule.

## CHARGE ORDERING AND VERTEX FUNCTIONS IN ARCS SCENARIO

The behavior of the charge ordering vertex corrections in the Fermi arcs scenario is presented in Figure 1. This should be compared with the analogous figure in the paper for the Fermi pockets scenario. Here there are no hot spots- and therefore no sharp features. Charge ordering in the arcs case (PRB 88, 064516 (2013)) focused on fluctuations at finite $\omega$, where the role of $\omega \neq 0$ and finite $\gamma$ (both are effectively "pairbreaking"), revealed signatures of the anti-nodal bandstructure.



\end{document}